\documentstyle[preprint,eqsecnum,aps]{revtex}

\def\eq{\begin{equation}}
\def\en{\end{equation}}
\begin{document}
\draft
\title{\bf One-dimensional pair-hopping and attractive Hubbard models:
A comparative study}
\author{Mathias van den Bossche* and Michel Caffarel**,\\
{*}Laboratoire CNRS-Physique th\'eorique des liquides,\\
Tour 16, 4 place Jussieu, 75252 Paris Cedex 05, France\\
{**}Laboratoire CNRS-Chimie Th\'eorique\\
Tour 22-23, 4 place Jussieu, 75252 Paris Cedex 05, France}
\date{\today}
\maketitle

\begin{abstract}      
The low-energy physics of the one-dimensional Pair-Hopping (PH) 
and attractive Hubbard models are expected to be similar.
Based on numerical calculations on small chains, several 
authors have recently challenged this idea and predicted the existence of 
a phase transition at half-filling and finite positive
coupling for the pair-hopping model.  
We re-examine the controversy by making systematic comparisons 
between numerical results obtained for the PH and attractive Hubbard models. 
To do so, we have calculated the Luttinger parameters 
(spin and charge velocities,
stiffnesses, etc...) of the two models
using both the Density Matrix Renormalization Group method for large systems
and Lancz\'os calculations with twisted boundary conditions for smaller systems.
Although most of our results confirm that both models are very similar we 
have found some important differences
in the spin properties for the small sizes considered by previous 
numerical studies (6-12 sites).
However, we show that these differences disappear at larger sizes (14-42 sites)
when sufficiently accurate eigenstates are considered.
Accordingly, our results strongly suggest that the ground-state phase 
transition previously found for small systems is a finite size artefact.
Interpreting our results within the framework of the Luttinger liquid theory,
we discuss the origin of the apparent contradiction between 
the predictions of the perturbative Renormalization group 
approach and numerical calculations at small sizes.

\end{abstract}
\pacs{PACS numbers:71.27.+a, 74.20.-z, 74.20.Mn}
\section{Introduction}
In this paper we are concerned with the pair-hopping (PH) model 
described by the Hamiltonian 
\eq
{\rm H = - t\sum_{<i,j>\sigma} [ c_{i\sigma}^+ c_{j\sigma} + h.c ] - V
\sum_{<i,j>}[c_{i\uparrow}^+
 c_{i\downarrow}^+ c_{j\downarrow}
c_{j\uparrow} + h.c.]
 }
\label{HamPH}
\en
\noindent where $c_{i\sigma}^+$ ($c_{i,\sigma}$) creates (destroys)
a fermion of spin $\sigma$ ($\sigma=\uparrow,\downarrow$) at lattice site $i$.
The first term is the usual kinetic energy term (tight-binding approximation),
the $V-$term allows spin-singlet pairs of electrons to hop from site to site. 
In what follows, we shall restrict our study to the case $V >0$\cite{note1}. 

There are a number of reasons which make this model interesting to study.
First, the pair-hopping model can be viewed as a phenomenological model 
to describe the dynamics of small size Cooper pairs. 
Since high-T$_c$ superconductors are known to display
such pairs, to study this model can be important 
to capture some of the physics of these materials.
Of course, when working with such a model nothing is said  about the 
nature of the underlying mechanism responsible for the tight binding 
of the pairs. 
Second, it can be shown that the pair-hopping term arises from Coulomb 
interaction at large negative U in the Hubbard model\cite{Emery,Nozieres}. 
Accordingly, 
the competition between the usual one-electron hopping and 
pair-hopping is related in some way to
the physics of the Hubbard model at strong coupling. Finally, 
understanding the physics resulting from all possible 
unusual interactions in 1D strongly correlated models is clearly a problem of 
central importance in solid state physics.

Very recently this model has led to some contradictory results. 
Using exact diagonalization calculations on small $1D$-chains
(up to L=10 sites, with periodic boundary conditions), 
Penson and Kolb claimed\cite{PK} that a phase transition
should occur at some finite critical value of the hopping parameter
V with V$_c$/t $\sim$ 1.4.
More precisely, they showed that a gap in the single particle spectrum
of the half-filled system opens up at that value.
They have also observed that the second derivative of the ground state energy
with respect to $V$ (a quantity similar to a
specific heat) has a local maximum at the transition which seems not to 
diverge.
This would indicate a phase transition with an essential singularity.
Very soon later,
Affleck and Marston,\cite{AM} making a renormalization-group analysis
with bosonization methods of the PH-model, showed that, in the continuum
limit (low-energy, long-distance physics), this model is
essentially equivalent, up to some irrelevant terms, to the
negative-$U$ Hubbard model,
the only important difference lying in the bare coupling constants.
Accordingly, they predicted that the transition in the pair-hopping
model must occur at V=0 just like in the
Hubbard model, the finite value observed in the numerical calculations 
for very small chains being attributed to a finite-size artefact.
A few years later, Hui and Doniach \cite{DH} presented some new numerical
calculations analysed with more sensitive tools than the standard
finite-size scaling analysis based on very small samples.
Using an eigenprojection
decomposition of the different order parameter operators involved and
also some calculations of the helicity modulus,
they found that the data seemed indeed to be compatible with
the existence of a phase transition at a finite value of V, thus in 
contradiction with the weak-coupling renormalization group results. 
They also presented some arguments on why the predictions of the 
renormalization group analysis of Affleck and Marston could be not valid.
Very recently, Bhattacharyya and Roy \cite{BR} have investigated the 
PH model using a real space renormalization group method. 
At small positive V they also found 
the existence of gapless phase (identified as a quasi-metallic 
phase dominated by short range superconducting correlations) which disappears 
at some finite value of the coupling.
Finally, Sikkema and Affleck\cite{AS} have presented some numerical
results for the one-particle gap as a function of V using the Density Matrix 
Renormalization Group (DMRG) method with open boundary conditions.
Using samples up to L=60, they concluded that
there is no spin-gap transition at a non-zero positive value of V and that the 
standard low-energy picture given by the perturbative Renormalization 
Group approach is valid. However, although we reach in this work essentially
the same conclusions (following a quite different route), 
we do not agree on the use of
open boundary conditions for this problem (See, section IV).

At the heart of the controversy is the question of knowing  whether
the long-distance, low-energy physics of the pair-hopping model
is different or not of that of the usual attractive Hubbard model. As we shall
see in the next section all standard approaches lead to the same conclusion: 
the low-energy sector of both Hamiltonians should be similar under the 
trivial correspondance $U =-2V$. At  
half-filling it is known (an exact result) that no phase transition 
at a nonzero value of U exists for the attractive Hubbard model. 
How can the PH model exhibit a different behavior? 
This should result from a highly non-trivial process involving non-trivial 
excitations. Note also that the exotic gapless phase is 
supposed to exist at arbitrary small value of the hopping parameter, a 
domain where the
high-energy degrees of freedom are not expected to play an important role.
In order to settle down the controversy we propose to make a systematic 
comparison of the physics of the Pair Hopping and attractive Hubbard models 
at low energy. 
To do so, we have calculated the spin and charge velocities of the two models
using both the Density Matrix Renormalization Group method with 
periodic boundary conditions for large systems 
and Lancz\'os calculations with twisted boundary conditions for smaller systems.
Our results show that there are some important differences in the finite-size
behavior of the two models. 
Using the framework of the Luttinger liquid 
we propose an interpretation of the origin of the 
controversy between the perturbative RG prediction and the numerical 
results for small chains presented up to now.

The paper is structured in the following way. In the second section, 
we briefly present the results of a number of approaches 
illustrating the very close similarity between the
attractive Hubbard model and 
the Pair Hopping model. 
In the following section, we present our numerical results 
using the Luttinger Liquid theory and the 
twisted boundary conditions method on both models for chains up to L=12 sites.
Then, using the DMRG method we generalize the results presented for 
small chains at some larger chains up to L=42 sites.
In the last section, we discuss the results
and comment on what we believe to be the origin of the controversy.
We conclude that: 1) both models are indeed equivalent at low-energy
in the thermodynamic limit and that there is no phase transition at finite V 
and  half-filling in the PH model
2) for small systems there exists a transient regime specific to the PH 
model and responsible for the unconventional behavior of this model.

\section{Pair Hopping and Attractive Hubbard models}

The Hamiltonian (\ref{HamPH}) for the Pair Hopping model describes a competition 
between the usual kinetic term (t-term) corresponding to single-electron 
hopping and a V-term corresponding to the hopping of spin-singlet
pairs, the range of both types of hopping being limited to nearest neighbors.
When V/t is large ($V>0$), the pair-hopping term dominates and the model becomes 
equivalent to spinless fermions (for an even number of electrons). The
ground state is massively paired and there is a gap of order V in the
one-particle spectrum (binding energy of the pairs). 
In the opposite limit, $V/t<<1$, the one-particle 
hopping dominates and the pairs tend to be destroyed. This type of competition 
is very similar to that encountered in the attractive Hubbard model 
described by the Hamiltonian:
\eq
{\rm H = - t\sum_{<i,j>\sigma} [ c_{i\sigma}^+ c_{j\sigma} + h.c ] +U 
\sum_{i}n_{i\uparrow}
 n_{i\downarrow}
 }
\en
Here also, we are in presence of a competition between a one-electron 
hopping and the formation of spin-singlet pairs. However, in contrast with 
the PH model, pairs have no intrinsic mobility (uncorrelated mobility 
via the t-term). The physics of the
attractive Hubbard model is well understood since this model admits an exact 
solution via the Bethe Ansatz technique. In particular, it is known that 
the effect of the on-site interaction is rather drastic: a 
gap in the one-particle spectrum opens up for any nonzero value of the 
interaction U (negative or positive) at half-filling. It is usually thought that 
a similar situation should occur in the PH model. This opinion is 
supported by the fact that standard approximate approaches applied to 
both Hamiltonians lead quite systematically to the same physics
at low-energy  under the trivial correspondance $U\leftrightarrow -2V$. 
However, as already emphasized, this idea has been recently 
challenged. This is the purpose of the next few sections to shed some light 
on this controversy. Here, we would like briefly illustrate by applying some 
standard methods why the correspondance between both models is usually 
taken for granted.

A first approach to consider is the Mean Field Approximation (MFA).
Defining the superconducting order parameter by
$\Pi=<gnd|c_{i\downarrow}c_{i\uparrow}|gnd>$, where $|gnd>$ denotes the
BCS-type ground state, we consider the quantum fluctuations around this
value and construct the approximate Mean-Field Hamiltonian by
keeping only the terms which are of first-order with respect to the
fluctuations. The following Hamiltonian is obtained
\eq
H_{MF}=\sum_{\bf k, \rm\sigma} \varepsilon(\bf k \rm) c_{\bf k\rm,\sigma}^+
 c_{\bf k\rm,\sigma} - 4V\Pi D\sum_{\bf k \rm}[ c_{\bf k\rm,\uparrow}^+c_{-\bf
k\rm,\downarrow}^+ + c_{\bf k\rm,\downarrow} c_{-\bf k\rm,\uparrow}]
+4V\Pi^2 DL^D,
\en
\noindent where $\varepsilon(\bf k \rm)=-4t \sum_{\mu=1}^D
\cos(\bf k.e\rm_\mu)$, $\bf e\rm_\mu$ being the unit vector in direction $\mu$,
and
D the dimension of space.
The main observation is that this Hamiltonian is identical to that obtained
in the case of the Hubbard model\cite{LM} with the substitution $U=-2V$.
Introducing the elementary excitations in the usual way, we can compute
the dependence of the gap $\Delta$ in energy of the system, we get:
$$
\Delta \sim t e^{-c\frac{t}{|V|}} \mbox{\hspace{1cm}for
$V\rightarrow 0,\;\;$ where $c$ is a positive constant}
$$
and
\noindent
\eq
\Delta \sim V \mbox{\hspace{1cm}for $V\rightarrow \infty$},
\en
\noindent Clearly, in this approach both models are equivalent and the gap opens 
up at $V/t=0$ with a standard behavior.

We have also considered the large-dimension limit
of the pair-hopping model. This recent approach can be seen as a sort of dynamical mean field theory.  
Although this limit may seem rather academic,
practical calculations have illustrated the fact that a great part
of the physics of low-dimensional systems is captured \cite{MV,KGK}. Once again,
in that approximation we have found that the equations reduce to
those of the corresponding attractive Hubbard model with $U=-2V$.
In fact, this is not really surprising since, because of the structure of 
the Fermi hypersurface in the limit of large dimensions,
the effects of the high-energy excitations which could 
be responsible for non-trivial processes are strongly suppressed.

As we shall see in Sec. V the Renormalization Group (RG) flows 
in the weak-coupling limit are also identical for the two 
models (Eq.(\ref{EqsRG})) with, here also, the same correspondance 
between couplings. This is only the initial values of the coupling 
constants which are model-dependent.

Finally, one can try to find out whether the PH-model has an exact solution 
via Bethe Ansatz. 
The essential step is to compute
the two particle S-matrix from the Schr\"odinger equation and then
to verify whether the S-matrix satisfies the Yang-Baxter (YB) condition.
Denoting by $A_{\sigma_1,\sigma_2}(p_1,p_2)$ the amplitude of the two-particle
wave function written in terms of a combination of plane waves,
defining as usual the two-particle S-matrix as follows
\begin{equation}
A_{\sigma_2,\sigma_1}(p_2,p_1)=\sum_{\sigma_1',\sigma_2'} S_{\sigma_2,\sigma_2'}
^{\sigma_1,\sigma_1'}(p_1,p_2)A_{\sigma_1',\sigma_2'}(p_1,p_2),
\end{equation}
forcing the wave function to obey the Schr\"{o}dinger equation, and
imposing the continuity condition of the wave function, we get
the following expression for the S-matrix:
$$\
S_{\sigma_2,\sigma_2'}
^{\sigma_1,\sigma_1'}(p_1,p_2)=
\frac{\sin{ap_1}-\sin{ap_2}}{\sin{ap_1}-\sin{ap_2}- i V\cos[a(p_1+
p_2)]}
\delta_{\sigma_1,\sigma_1'}\delta_{\sigma_2,\sigma_2'}
$$
\eq
-\frac{ i V \cos[a(p_1+p_2)]}{\sin{ap_1}-\sin{ap_2}
- i V\cos[a(p_1+ p_2)]}
\delta_{\sigma_1,\sigma_2'}\delta_{\sigma_2,\sigma_1'}
\label{Smatrix}
\en
It is easy to verify that the S-matrix just given does not
satisfy the Yang Baxter condition \cite{YS}. Now, the important
point is that the S-matrix (\ref{Smatrix}) is identical to that of 
the Hubbard model with the substitution 
$ U\rightarrow -2 V \cos{[a(p_1+p_2)]}$.
The lattice spacing $a$ gives a natural high-energy cut-off, $1/a$, in
the problem. In the low-energy regime, i.e. $p_i << 1/a$,
both approaches lead to the same equations and the two models
related by $U=-2V$ should be equivalent.

To summarize, mean-field approximation, large-D
limit,  weak-coupling  renormalization group and Bethe Ansatz approaches 
indicate that the PH-model and the $U=-2V$ attractive Hubbard model should 
be equivalent in the low-energy regime.

\section{Luttinger liquid behavior: an exact diagonalization study on small systems}

In this part we are interested in evaluating the Luttinger liquid parameters 
for both the Pair Hopping and attractive Hubbard models. As is well-known 
the long distance behavior of one-dimensional gapless fermion systems can 
be studied by making use of the concept of "Luttinger liquid". 
Within the framework of this theory the low-energy properties are given by 
an effective Luttinger model describing collective spin and 
charge density oscillations. The general form of the 
effective Hamiltonian can be obtained by writing the 1D fermion model 
in momentum space, restricting excitations and interactions to lie close to 
the Fermi surface, and looking for the important processes.
As well known only four 
processes survive (in the Renormalization Group sense): 
one describing backward scattering of oppositely 
moving electrons with coupling\footnote{Notations are those of references 
\cite{Sol,Voit}} $g_1$, one describing forward scattering of oppositely 
moving electrons with coupling $g_2$, one describing Umklapp scattering 
with coupling $g_3$ and, finally, one describing forward scattering 
of electrons moving 
in the same direction with coupling $g_4$. Taking the continuum limit of 
the fermion Hamiltonian and, then, bosonizing the fermi fields, one gets:
\eq
H_b=H_\rho+H_\sigma + H_1+H_3
\en
where $H_\nu (\nu =\rho,\sigma)$ are two free bose Hamiltonians describing 
the spin ($\nu=\sigma$) and charge ($\nu=\rho)$ collective excitations:
\eq
H_\nu=\int dX [\frac{u_\nu}{2\pi K_\nu}(\partial_X \phi_\nu)^2+
\frac{u_\nu \pi K_\nu}{2} \Pi_\nu^2]
\en
and $H_1$ and $H_3$ are the terms corresponding to the
backward and Umklapp scattering contributions, respectively
\eq
H_1=\frac{2g_1}{(2\pi\alpha)^2}\int dX cos(\sqrt8\phi_\sigma)
\label{h1}
\en
and
\eq
H_3=\frac{2g_3}{(2\pi\alpha)^2}\int dX cos(\sqrt8\phi_\rho).
\label{h3}
\en
Here, $\phi_\rho$ (resp. $\phi_\sigma$) is the bose field describing the 
charge (resp. spin) excitations, and $\Pi_\rho$ (resp. $\Pi_\sigma$) is its 
canonical conjugated field. 
The coefficients $u_\rho$ (resp. $u_\sigma$) are the charge (resp. spin) 
excitation velocities, and the parameters $K_\rho$ and $K_\sigma$ are
some constants which can be shown to be related to the (non-universal) 
exponents of the power-law behavior of the correlation functions. 
In Eqs.(\ref{h1}) and (\ref{h3})
$\alpha$ is a short-distance cut-off$^{\cite{Voit}}$. 

In the free-fermion case, $K_\rho = K_\sigma = 1$ and $u_\rho= u_\sigma = v_F 
= 2t sin(\frac{\pi}{2}n)$, where $n=N/L$ is the electron density. 
When interactions are switched on, the $u$'s and the $K$'s parameters are 
renormalized. In particular, the two velocities become different, 
charge and spin excitations do not propagate at the same speed. This phenomena 
is known as the spin-charge separation in one-dimensional systems.
All the details concerning the Luttinger liquid theory can be found, e.g., 
in Refs.\cite{Voit,Sol} and references therein.

In order to compute numerically the Luttinger coefficients, we have used their
expressions in terms of spin and charge compressibilities and 
stiffnesses of the system. More precisely, for the charge degrees of freedom we
have
\eq
\frac{1}{\kappa}=\frac{\pi}{2}\frac{u_\rho}{K_\rho}
\;\;\;\;\;\;\;\;\;\;\;\;\;\;\;\;\;
D_\rho=2u_\rho K_\rho
\en
\noindent where $\kappa$ is the compressibility of the system and 
$D_\rho$ is the charge siffness, and for the spin degrees: 
\eq
\frac{1}{\chi}=\frac{\pi}{2}\frac{u_\sigma}{K_\sigma}
\;\;\;\;\;\;\;\;\;\;\;\;\;\;\;\;\;
D_\sigma=2u_\sigma K_\sigma
\en
\noindent where $\chi$ is the spin susceptibility of the system and 
$D_\sigma$, the spin stiffness.
These quantities can be computed from the spectrum of the system
by using the relation$^{\cite{SSGV}}$:
\eq
D_\nu=\left .
\pi \frac{\partial^2 E_0}{\partial \varphi_\nu^2}
\right|_{\varphi_\nu=0}
\en
\noindent where $\varphi_\rho$ is a charge twist in the system, (i.e. 
the system has twisted boundary conditions such as 
$c^+_{j+L,\sigma}= exp(i\varphi_\rho)c^+_{j,\sigma}$), 
and $\varphi_\sigma$ is a spin twist in the system, 
(i.e. $c^+_{j+L,\sigma}= exp(i\sigma \varphi_\sigma)c^+ _{j,\sigma}$), and 
\eq
\frac{1}{\kappa}=\frac{1}{2L}\frac{\partial^2 E_0}{\partial n^2}, \;\;\;\;\;\;\;\;\;\;
\frac{1}{\chi}=\frac{1}{2L}\frac{\partial^2 E_0}{\partial s_Z^2} 
\en
\noindent with $n=(N_\uparrow+N_\downarrow)/L$ and 
$s_Z=(N_\uparrow-N_\downarrow)/L$. By computing these quantities for
different values of
the interaction, we can deduce the behavior of the 
Luttinger parameters, $u_\nu$ and $K_\nu$ as a function of the coupling strength.

We have applied this approach on systems of sizes ranging from $L=4$ to 
$L=12$. The ground-state energies have been calculated using a 
standard Lancz\'os procedure. The results are presented in figures 1-5. 
Each figure shows the variation of the corresponding Luttinger coefficient
as a function of the interaction, both for the attractive Hubbard 
model (squares) and for the Pair-Hopping model (crosses). 
Figure 1 gives the variation of the charge velocity, $u_\rho$, as a function of
U or V. At small coupling both curves are linear with a very good accuracy.
More precisely, we find $u_\rho \sim 2 + V/2 $ and $u_\rho \sim 2 + U/4$,
for the pair-hopping and Hubbard models, respectively. 
For stronger couplings, small corrections to linearity show up.
Both behaviors are typical of a regime with no charge gap.
As we shall see later, these results are
in perfect quantitative agreement with the prediction of the Luttinger liquid 
theory (Eqs.(\ref{init}) and (\ref{Eqsurho})). 
Data for the spin velocities are rather different.
As can be seen in Figure 2 two distinct behaviors for the spin 
velocity are obtained. In the case of the attractive Hubbard 
model $u_\sigma$ decreases uniformly from the free fermion value to zero
at large coupling. In contrast, a maximum around $V=0.55t$ is found 
for the pair hopping model. Both models recover a similar behavior 
between approximately $V=1.$ and $V=1.5$. 
Note that the transition value observed 
in Refs. \cite{PK,DH} lies within this interval. We shall discuss 
further this important difference of behavior for $u_\sigma$ in Sec. V. 
Figures 3 and 4 demonstrate that the constants
$K_\rho$ and $K_\sigma$ behave essentially the same way in both models. 
As already mentioned, in the Luttinger liquid theory these 
constants are related to the exponents of the power-law behavior of 
correlations functions. Accordingly, this common behavior would suggest 
that both models have the same phases.
In figure 5, the behavior of the spin stiffness of the pair hopping model
as a function of the size is displayed. A very interesting feature is 
that this quantity can be exactly computed for the Hubbard model.
The formula is \cite{SM}:

$$
D_\sigma(L)=(-1)^{L/2+1}L^{1/2}D(U)e^{-\frac{L}{\xi_\sigma(U)}}
$$

\noindent with
$$
\xi^{-1}_\sigma(U)=\frac{4}{U}\int_1^\infty dy\frac{ln{(y+\sqrt{y^2-1}})}{cosh{(2\pi y/U)}}
$$
\noindent and
\eq
D(U)\sim \sqrt{(2/\pi\xi_\sigma)}\;\;\;for\;\; U\rightarrow 0\;\;\;\;\; {\rm and}\;\;\;\;\;
D(U)\sim 0.147376\;U \;\;\;{\rm for}\;\; U\rightarrow \infty
\en
\noindent This function is plotted in figure 5, for $U/t=-2$, 
with the corresponding quantity for the pair hopping model, at $V/t=1$. 
The similarity between the two curves is striking. 
In the case of the Hubbard model, the oscillations around zero are related to
the existence of a gap in the spin spectrum. In the case of a gapless mode, the 
corresponding curve is smooth and never changes sign. Accordingly, we have here 
a strong evidence in favor of the existence of a spin gap in the pair hopping 
model.

At this point, our results are contradictory. On one hand most of the results 
indicate that both models are quite similar 
(behavior of $u_\rho$, $K_\nu$'s, and spin stiffnesses). 
On the other hand, the spin velocities at small sizes for both models 
display a different behavior. A closer look on spin degrees of freedom at 
larger sizes is therefore necessary.

\section{Luttinger liquid behavior: a DMRG study for larger systems}

Conformal field theory (CFT) is a powerful theory to describe
the physics of $1D$ quantum (or $2D$ statistical) critical systems.
Once conformal invariance is supposed, CFT provides a general framework 
relating finite-size scaling of physical quantities to thermodynamic 
properties\cite{Cardy,Woynarovich,FK}. In this work we shall essentially 
compare our data for excitation gaps with the predictions of CFT. This will 
allow us to check whether or not our data are compatible with the 
existence of a critical regime for the pair-hopping model. 
Denoting $\nu$ the gapless excitation 
under consideration and $u_\nu$ the velocity of the corresponding critical 
mode, the finite-size scaling (FSS) expression of the excitation gap 
$\Delta_\nu$ predicted by CFT is
\eq
\label{mat1}
\Delta_\nu=\frac{2\pi u_\nu}{L}
\en
where L is the system size.
\noindent For a finite system at a given filling, the spin gap is defined as
$$\Delta_\sigma=E_0(N_\uparrow+1,N_\downarrow-1)-E_0(N_\uparrow,N_\downarrow)$$
\noindent where $N_\sigma$ is the number 
of $\sigma$-spin electrons. Physically, it gives the change in ground 
state energy produced when flipping one spin, the charge number 
being kept fixed. 

In order to calculate the spin gaps we have used the Density Matrix 
Renormalization Group (DMRG) method \cite{White}. DMRG is a powerful 
technique to compute low-energy properties of quantum lattice systems.
This method has been applied with success to several problems including the 
spin-1/2 Heisenberg chains \cite{White}, the spin-1 chains \cite{WH}, 
the one-dimensional Kondo insulator \cite{YW}, the two-chain Hubbard model 
\cite{NWS}, etc... 
The results obtained are very accurate and the method
allows to treat systems of sizes a few times larger than those 
accessible with exact diagonalization techniques. Essentially, DMRG 
is a real-space numerical Renormalization Group procedure. It differs 
from standard approaches in the way that states of individual 
blocks are chosen. Instead of keeping the lowest eigenstates of the 
block considered as isolated from the outside world, the kept states 
are the most probable eigenstates of the density matrix associated with 
the block considered as a part of the whole system.
It is easy to show that doing this is equivalent to construct the 
most accurate representation of the complete state of the system: 
block plus rest of the system. 
For a detailed and very clear presentation of the 
method the reader is referred to \cite{White}. There exist different 
ways of choosing the configuration of blocks used for the density matrix 
calculations. In particular, this choice will depend on the 
type of boundary conditions 
used. Here, all calculations have been done by using periodic 
boundary conditions (PBC). We have chosen the superblock configuration 
$B_l\bullet B_l^R\bullet$ with $B^\prime_{l+1}= B_l\bullet$ as proposed 
in \cite{White} for PBC. $B_l$ represents a block consisting of l sites, 
$B_l^R$ is the reflected block (right interchanged with left), and 
$\bullet$ represents a single site. All notations are those of 
Ref. \cite{White}. 
In what follows we shall denote M the number of 
eigenstates of the density matrix kept. 

Very recently, Sikkema and Affleck have presented DMRG calculations for the 
pair-hopping model\cite{AS}. 
Their calculations have been performed using open 
boundary conditions (OBC) instead of PBC (calculations with OBC are less 
demanding than PBC). Using OBC can introduce 
important boundary effects.
Accordingly, OBC can be used only 
when the correlation length is known to be finite and when $L>>\xi$ (no 
boundary effects, no finite bias for the excitation gaps calculated as 
difference of energies of order O(L)). Therefore, when searching for 
a hypothetical gapless phase it is essential to use periodic boundary 
conditions. In the regime of small $\xi$ where the use of OBC is justified 
Sikkema and Affleck have shown that their data are consistent with the 
prediction of the standard perturbative RG flow. In this work we shall use PBC 
in a regime where the correlation lengths are large.

To begin with we present some DMRG calculations for the attractive 
Hubbard model. The value of the Coulomb interaction, U=-1.1, has been 
chosen to correspond to V=-U/2=0.55, the value for which the spin 
velocity of the PH model is maximum, see Fig. 2. 
Since the Hubbard model admits an exact solution 
our results can be compared to the exact values obtained by solving the 
Lieb-Wu equations \cite{Lieb}. Inset of Fig. 6 shows how the 
DMRG spin gap $\Delta_\sigma$ converges to the exact value 
$\Delta_\sigma=0.9297..$ for a chain of 14 sites as a function of M, the number 
of states kept. Here, M ranges from M=16 to M=112. Clearly, 
the convergence of the DMRG values is quite good. In addition, this curve 
provides a useful check of 
the validity of our code. The main plot displays the variation 
of the spin gap as a function of 1/L. The studied sizes are ranging from 
L=6 to L=42.
We did not consider the system sizes corresponding 
to a multiple of 4 since, in this case, the ground-state is degenerate, thus 
causing a strong boundary frustration effect (which, of course, disappears 
in the $L \rightarrow \infty$ limit). For each size, we plot the value 
of the DMRG spin gap for a number of kept states M=96, M=112, and 
M=$\infty$ (exact Lieb-Wu value). Let us first consider the exact solution.
Looking at the $L \rightarrow \infty$ limit, we observed a very small gap 
as expected. In this regime the systems considered (L=6-42) are 
in an effective quasi-critical regime with a spectrum 
structure remaining close to the conformal tower structure. This allows 
to write the following ansatz:
\eq
\label{mat2}
\Delta_\sigma (L)=\Delta_\sigma^\infty+\frac{2\pi u_\sigma}{L}
\en
valid in the regime $a<<L<<\xi$, and where $u_\sigma$ should be 
considered as an effective spin velocity. The results obtained are in excellent 
agreement with the behavior predicted by formula (\ref{mat2}) with 
a spin velocity very close of the free value. In addition, for
small systems (L=6,10) the spin velocity obtained from the slope of the 
spin gap is in very good agreement with the value obtained in the 
preceding section (within 1.5\%) based on a completely independent evaluation.

Let us now consider the DMRG results. We have observed that, for large enough 
values of M, the linear behavior of the spin gap as a function of 
1/L is recovered. In fig.6 we show typical results for M=96 and M=112. The 
value of the spin velocity obtained from different M are displayed in figure 8
and are slightly smaller than the free value of 2. These results are 
consistent with a convergence to the exact value 
at large M. However, 
it is not possible from DMRG results to get an accurate estimate of the 
value of the gap itself. Indeed, although we have 
a good convergence of the results for a given size as a function of M (see 
inset of Fig. 6),
the extrapolated value of the gap using different sizes is a very sensitive 
quantity. In fact, it is not reasonable to discriminate between a small 
but finite gap and a strictly vanishing gap. We clearly see on Fig. 6 that the 
extrapolated gap is not at all converged as a function of M. In order 
to get converged values we would need much larger values of M which are
clearly beyond of reach of present computers. 

In figure 7 we present DMRG calculations for the Pair Hopping model at 
V=0.55. Results of the spin gap as a function of 1/L are presented for 
M=96, 112, and M=144. Here again we clearly see a quasi-critical regime 
very well described by formula (\ref{mat2}). As already emphasized 
for the attractive 
Hubbard model, the accessible values of M do not allow a direct conclusion 
on the existence or not of a finite spin gap. However, the data provide an estimate 
of the effective spin velocity via the slope of the curves. The 
spin velocities obtained are plotted for M=84,96,112, and 144 in Fig. 8. 
It is remarkable that the results are rather different for both models. 
As already noticed, for the Hubbard model the values of $u_\sigma$
are slowly varying and always smaller than the free fermion value. 
In contrast, for the PH model $u_\sigma$ is quite 
important for small values of M and decreases uniformly for increasing M. 
Only when large enough values of M are used, spin velocities of both models 
become comparable. We shall comment more on this point in the next section.

\section{Discussion}

Let us summarize the results obtained. For small sizes (L=4-12) we have 
computed the Luttinger parameters $u_\rho,u_\sigma,K_\rho,K_\sigma,D_\rho$,
and $D_\sigma$ as a function of the interaction for both the attractive 
Hubbard and Pair-Hopping models. Regarding charge degrees of freedom 
all results for both models are consistent with the existence of a 
vanishing charge gap for arbitrary values of the interaction and
with the fact
that the low-energy charge sectors of both models are very similar.
These results are in agreement with the conclusions of previous 
works. 

Now, regarding spin degrees of freedom the situation is not so clear.
For small sizes our results show that parameters $K_\sigma$ and 
$D_\sigma$ for both models are almost identical (See Figs. 4 and 5). 
In particular, in the case of the PH model we clearly see
the oscillations of $D_\sigma$ around zero as a function of the size L, 
a behavior which is usually interpreted as resulting from 
the existence of a gap. However, data for the spin velocity of the PH model 
do not display the expected behavior of a system with a gap.
In contrast with the case of the attractive Hubbard model 
for which $u_\sigma$ decreases 
uniformly from the free fermion value to zero at large coupling (a typical 
behavior for a finite system with a finite gap in the thermodynamic limit),  
we have observed a clear enhancement 
of $u_\sigma$ when the pair-hopping term is switched on. 
At V $\sim$ 0.55t the spin velocity of the PH model reaches a maximum 
and, then, decreases to zero. A similar behavior is recovered for 
both models at approximately V $>$ 1.5. In order to understand whether this 
surprising result has something to do with the existence of a gapless 
phase we have computed the spin gaps for larger systems 
using a DMRG approach with periodic boundary conditions. 
Extracting from the spin gaps some effective spin velocity (meaningful only 
when correlation lengths are much larger than lattice sizes) we have, here 
also, systematically obtained larger spin velocities for the PH model.
In contrast, in the case of the Hubbard model the spin velocities are 
rather constant and are close to the free fermion value at small coupling.
However, a remarkable result is that the abnormally large values of  
$u_\sigma$ for the PH model tend to disappear when sufficiently accurate 
representations of the ground-state of the system are considered (large number 
of kept states for the density matrix). Accordingly, our results 
are consistent with the fact that the unconventional behavior of spin 
excitations of the PH model is a transient effect specific to this model.

Now, it is quite interesting to discuss our results within the Renormalization 
Group framework.  As discussed very 
recently by Sikkema and Affleck, contradictory results have been reported 
from the RG analyses of the phase diagram of the PH model. Using standard 
notations (see, Ref.\cite{Sol}), to cubic order, the RG equations for the four 
coupling constants of the continuum-limit Hamiltonian are:
$$
-\frac{dg_s}{dl} = g_s^2 + \frac{1}{2}(g_s+g_4) g_s^2
$$
$$
-\frac{dg_\rho}{dl} = g_3^2 + \frac{1}{2}(g_\rho-g_4) g_3^2
$$
$$
-\frac{dg_3}{dl} = g_\rho g_3 + \frac{1}{4}(g_\rho^2+g_3^2-2 g_\rho g_4) g_3
$$
\eq
-\frac{dg_4}{dl} =\frac{3}{4}(g_\rho g_3^2 - g_s^3)
\label{EqsRG}
\en
where $l=-\log{\Lambda}$, $\Lambda$ being the ultraviolet cutoff.
It is important to emphasize that these equations are identical for both models. 
The only difference lies in the 
initial values of the coupling constants. To the lowest-order weak-coupling 
limit the initial values are:
$$
{\rm v_F=2t \;\; g_\rho=-g_s=g_3=g_4= 2V/\pi v_F \;\; Pair \; \;Hopping \;\;model}
$$
\eq
{\rm v_F=2t \;\; g_s=-g_\rho=g_3=g_4= U/\pi v_F \;\; Hubbard \;\; model}.
\label{init}
\en
$O(V^2)$ corrections are given in Refs. \cite{AM} and 
\cite{DH}.
When solving the RG equations, a standard approach consists in considering that, 
$g_4$ simply shifts the spin and charge velocities according to:
\eq
u_\rho = v_F (1 + g_4/2)
\label{Eqsurho}
\en
\eq
u_\sigma = v_F (1 - g_4/2)
\label{Eqsusig}
\en
and then can be dropped from the RG equations. Doing this and using the initial 
conditions Affleck and Marston have remarked that $g_s= 0$ is not a stable fixed 
point and that starting with $g_s < 0$ (V $>$ 0) then $g_s$ flows 
to strong coupling, thus indicating the opening of a gap in the spin excitations.
In contrast, Hui and Doniach have kept the $g_4$ constant in the RG equations and 
integrated them using the initial condition at order $O(V^2)$. By doing this 
they obtained that $g_s=0$ becomes a stable fixed point provided that 
$g_4 < -2$.
For $ 0 < V/t < 1$ the fixed point was obtained with $g_4 \sim -2.5$. This 
new phase was interpreted as having no gap for single-particles and spin 
excitations. For a full discussion of the controversy the reader is referred 
to Ref.\cite{AS}. However, keeping or not the coupling constant $g_4$  
in the RG equations, it is clear that it is difficult to 
draw firm conclusions using weak-coupling RG equations in a strong coupling 
regime (fixed point with $g_4 \sim -2.5$). Nonpertubative results are essential 
to support any reasonable scenario. Let us discuss our numerical results 
from that point of view. Figure 1 shows very clearly that the charge velocity 
for both models follows exactly the behavior predicted by Eqs.(\ref{init})
and (\ref{Eqsurho}) with 
the correct slope. The charge degrees of freedom are gapless and the effect of 
the coupling constant $g_4$ is to renormalize the charge velocity. 
Figure 2 for $u_\sigma$ for small sizes is consistent with the fact 
that a spin gap exists for the attractive Hubbard model. The behavior of 
$u_\sigma$ is not linear at small U as would be the case for a
critical system. In addition, $u_\sigma$ decreases uniformly as a function 
of U. In contrast, as already pointed out we have found a different behavior 
for the PH model. At small sizes (L=6-10) and small coupling, $u_\sigma$ 
is larger than the free fermion value. This is also true for larger 
systems (L=14-42) when approximate ground-state wave functions are 
considered. From Eqs.(\ref{Eqsusig}) we can view this regime as 
corresponding to a situation where the effective constant $g_4$ starts to 
renormalize to negative values. In this situation the system appears to 
be attracted by a fixed point similar to the one discused by Hui and 
Doniach. However, as discussed before this is only a transient 
regime. When the low-lying eigenstates are sufficiently well described 
(large mumber of kept states in DMRG) the high-energy components responsible 
for this unconventional behavior are 
removed and the standard low-energy behavior is recovered. 
We believe that this very specific behavior of the PH model
is at the origin of the unconventional results obtained for sizes L=4,12 in
previous numerical works (Refs. \cite{PK},\cite{DH}, and \cite{BR}).

\section*{Acknowledgments}

We would like to acknowledge helpful discussions with 
P. Azaria, B. Dou\c cot, T. Giamarchi,
P. Lecheminant, C. Lhuillier, K.A. Penson, and C. Sire.
Part of computations have been done using an allocation of computer time from the "Institut du 
D\'eveloppement et des Ressources en Informatique Scientifique" (IDRIS, Orsay).

\newpage
\section*{Figure Captions}

\noindent
{\bf Fig. 1} Charge velocity $u_\rho$ as a function of the coupling. Crosses:
Pair-Hopping model, squares: Hubbard model. Lancz\`os calculations with 
twisted boundary conditions. Chains of sizes up to 12 sites.\\
\ \\
{\bf Fig. 2} Spin velocity $u_\sigma$ as a function of the coupling. Crosses:
Pair-Hopping model, squares: Hubbard model.
Lancz\`os calculations with
twisted boundary conditions. Chains of sizes up to 12 sites.\\
\ \\
{\bf Fig. 3} Charge parameter $K_\rho$ as a function of the coupling. Crosses:
Pair-Hopping model, squares: Hubbard model.
Lancz\`os calculations with
twisted boundary conditions. Chains of sizes up to 12 sites.\\
\ \\
{\bf Fig. 4} Spin exponent $K_\sigma$ 
as a function of the coupling. Crosses: Pair-Hopping model, 
squares: Hubbard model. Lancz\`os calculations with
twisted boundary conditions. Chains of sizes up to 12 sites.\\
\ \\
{\bf Fig. 5} Spin stiffness $D_\sigma$ 
as a function of the coupling. Crosses: Pair-Hopping model, 
squares: Hubbard model. Lancz\`os calculations with
twisted boundary conditions. Chains of sizes up to 12 sites.
The dotted line is just a guide to the eye.\\
\ \\
{\bf Fig. 6} Spin gap vs. inverse of the system size for the attractive Hubbard model 
at $U/t=-1.1$
using DMRG with periodic boundary conditions 
for different values of M (see text). Inset shows the convergence of 
the spin gap as a function of M at L=14 sites. The value at $1/M=0$ is 
the exact value calculated by solving the Lieb-Wu equations.\\
\ \\
{\bf Fig. 7} Spin gap vs. inverse of the system size for the Pair Hopping model
at $V/t=0.55$
using DMRG with periodic boundary conditions 
for different values of M (see text).\\
\ \\
{\bf Fig. 8} Spin velocity $u_\sigma$ computed from DMRG data as a function 
of 1/M.
\ \\
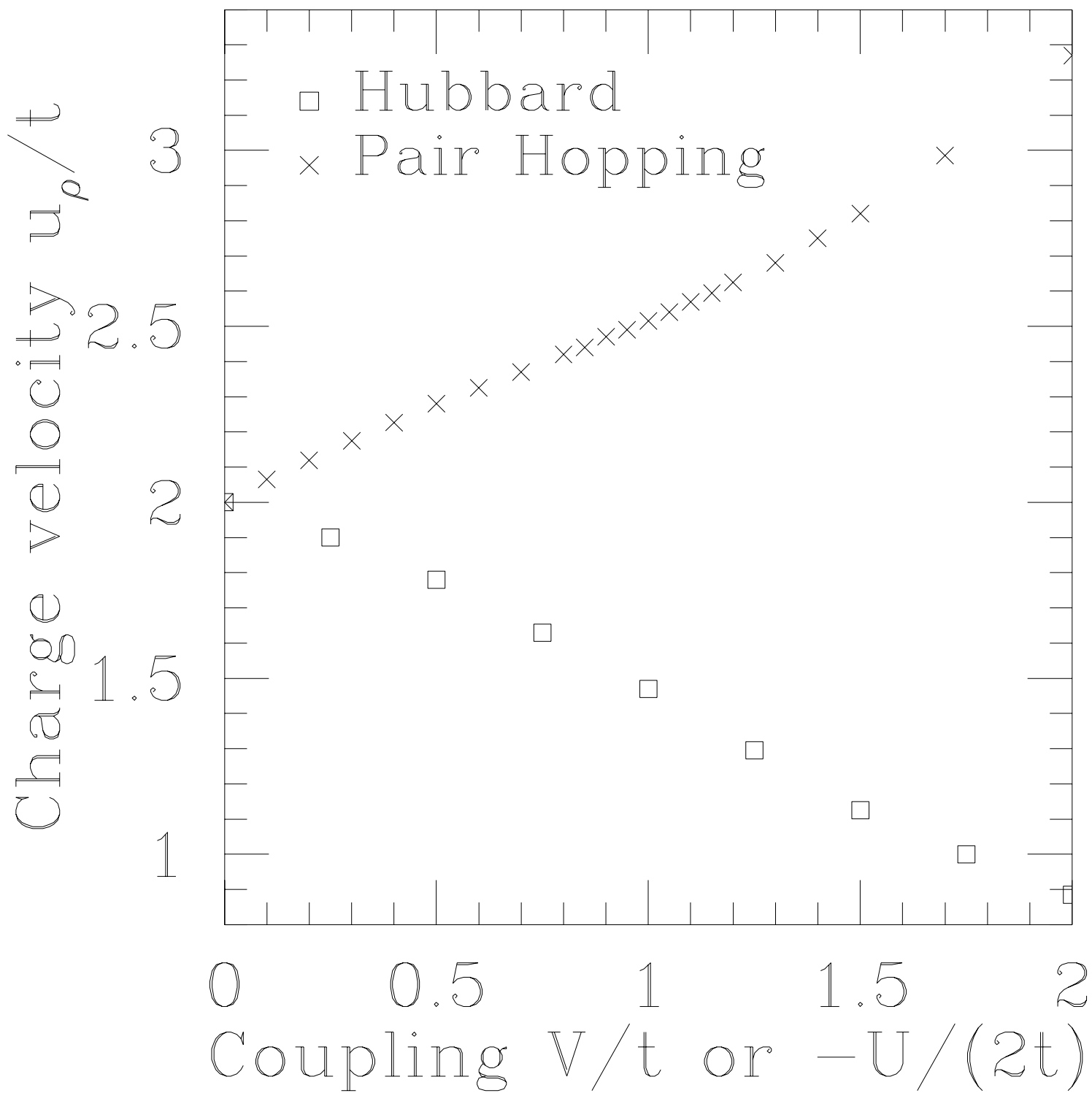
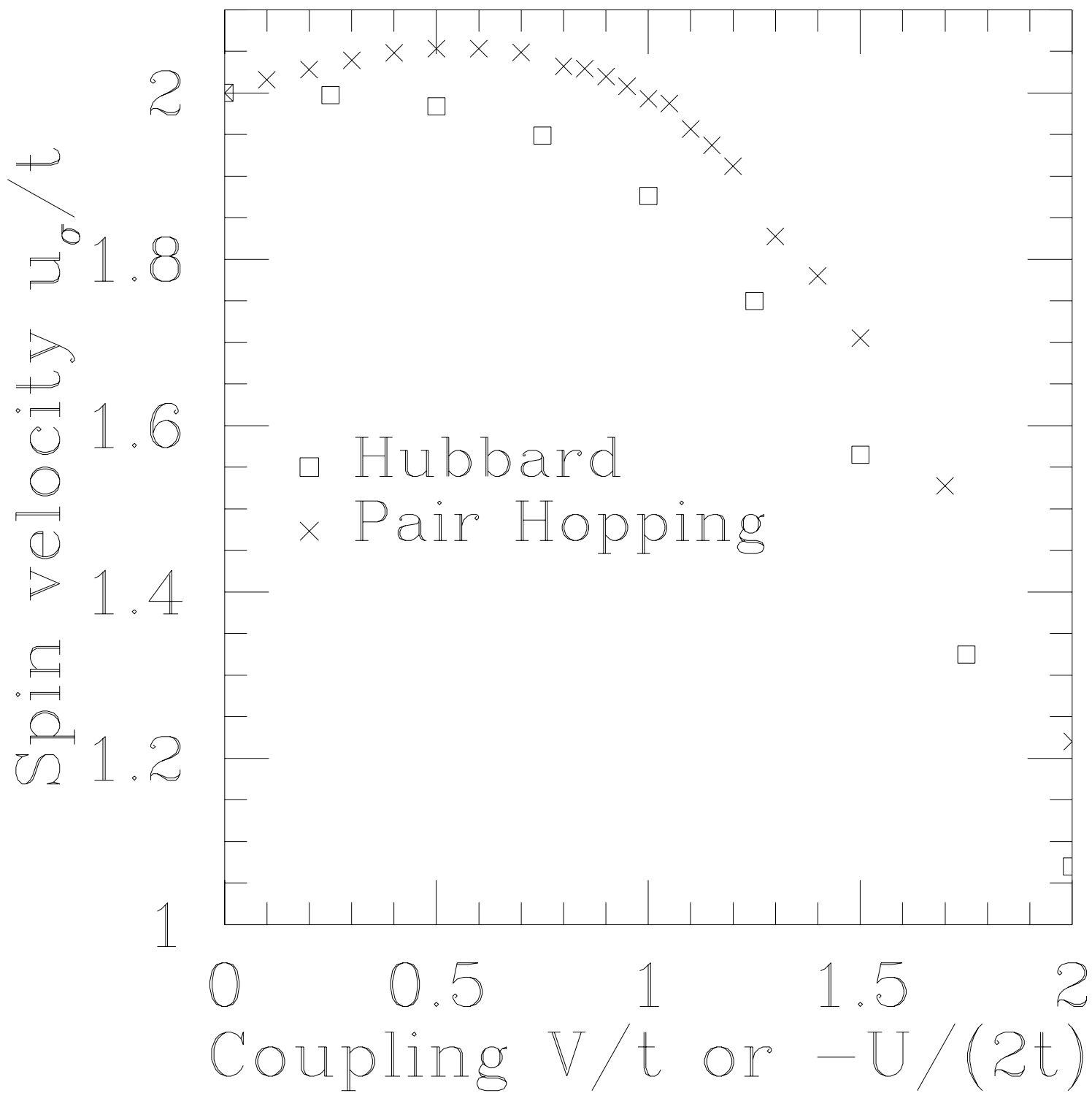
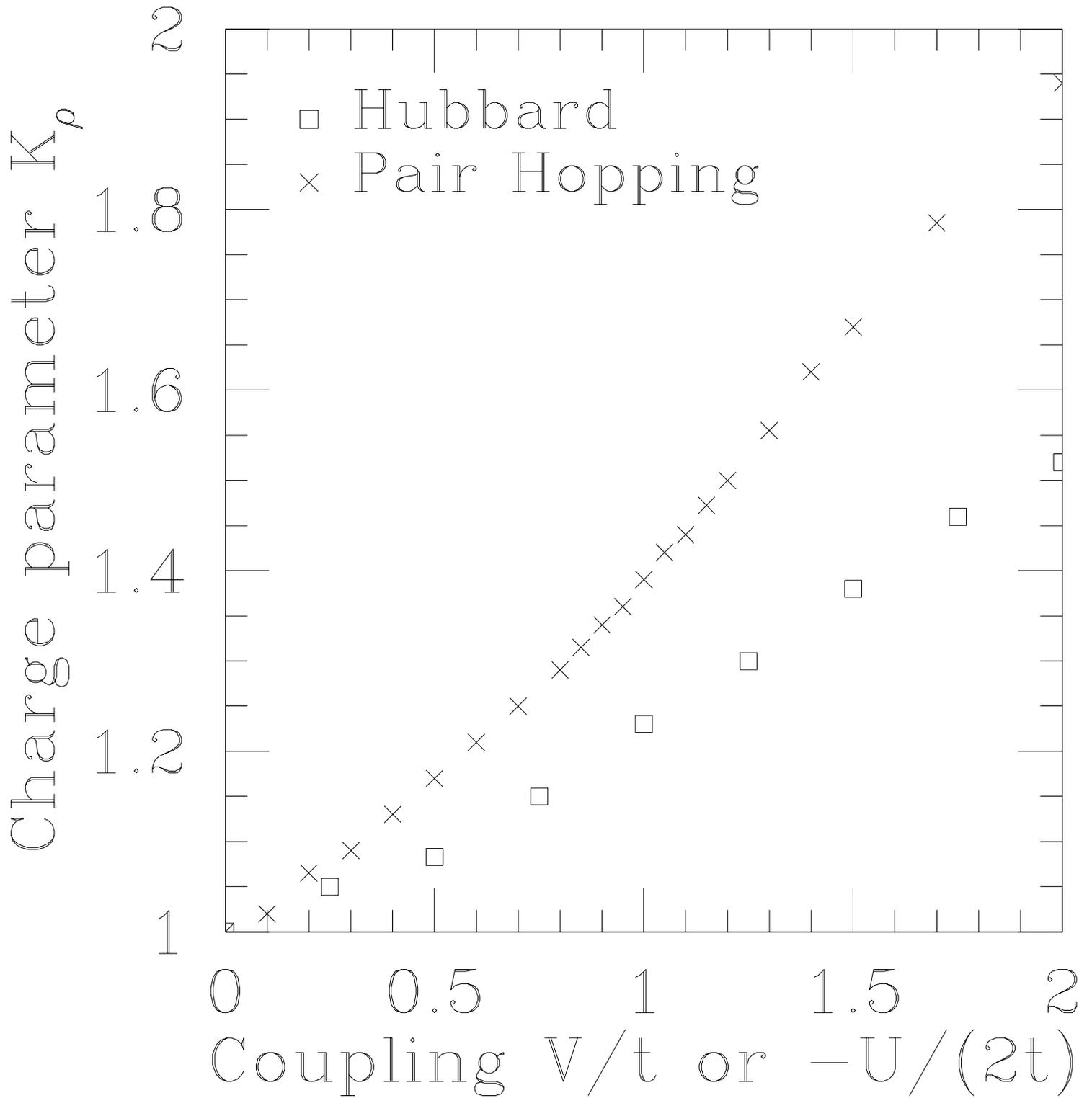
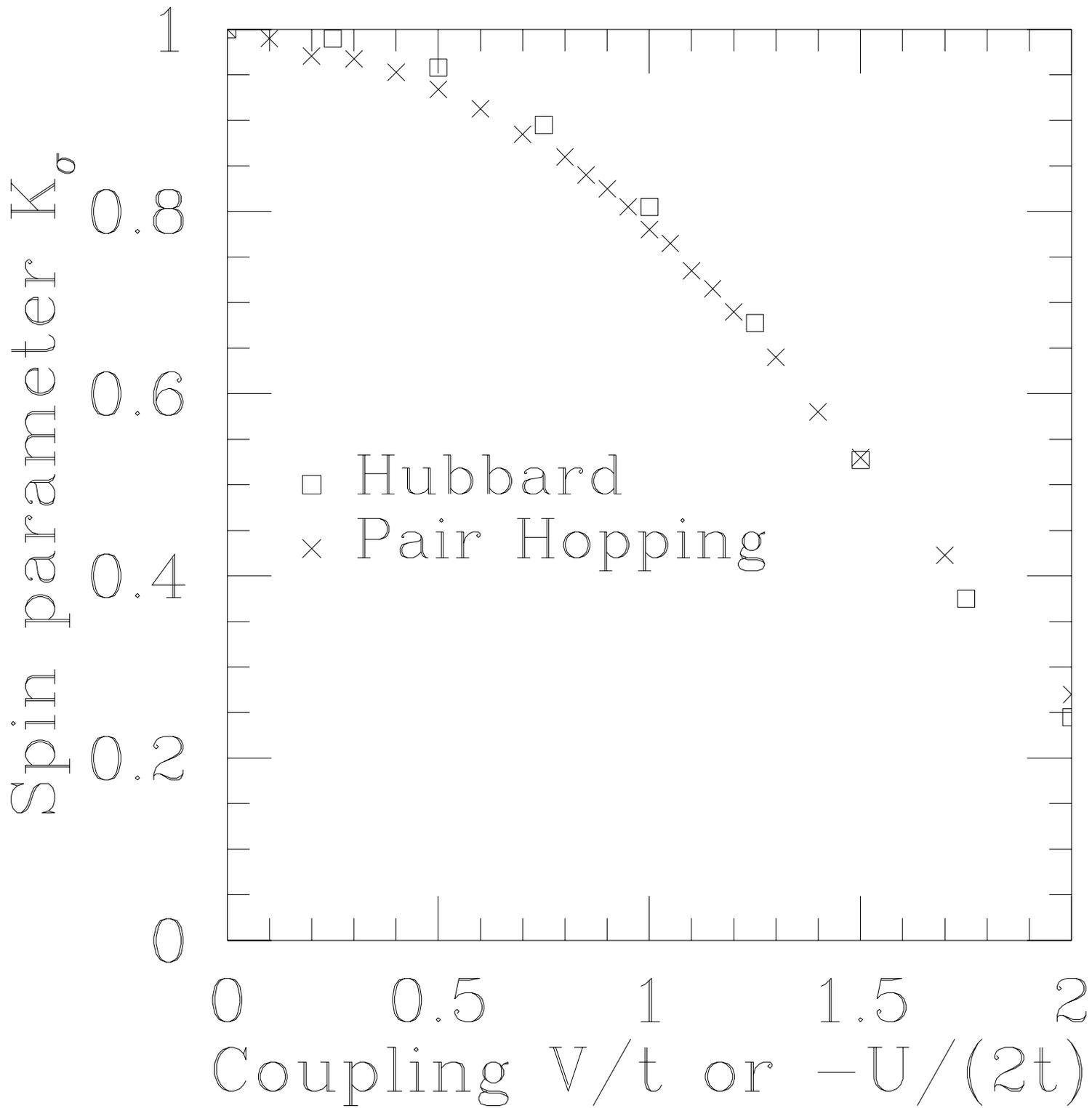
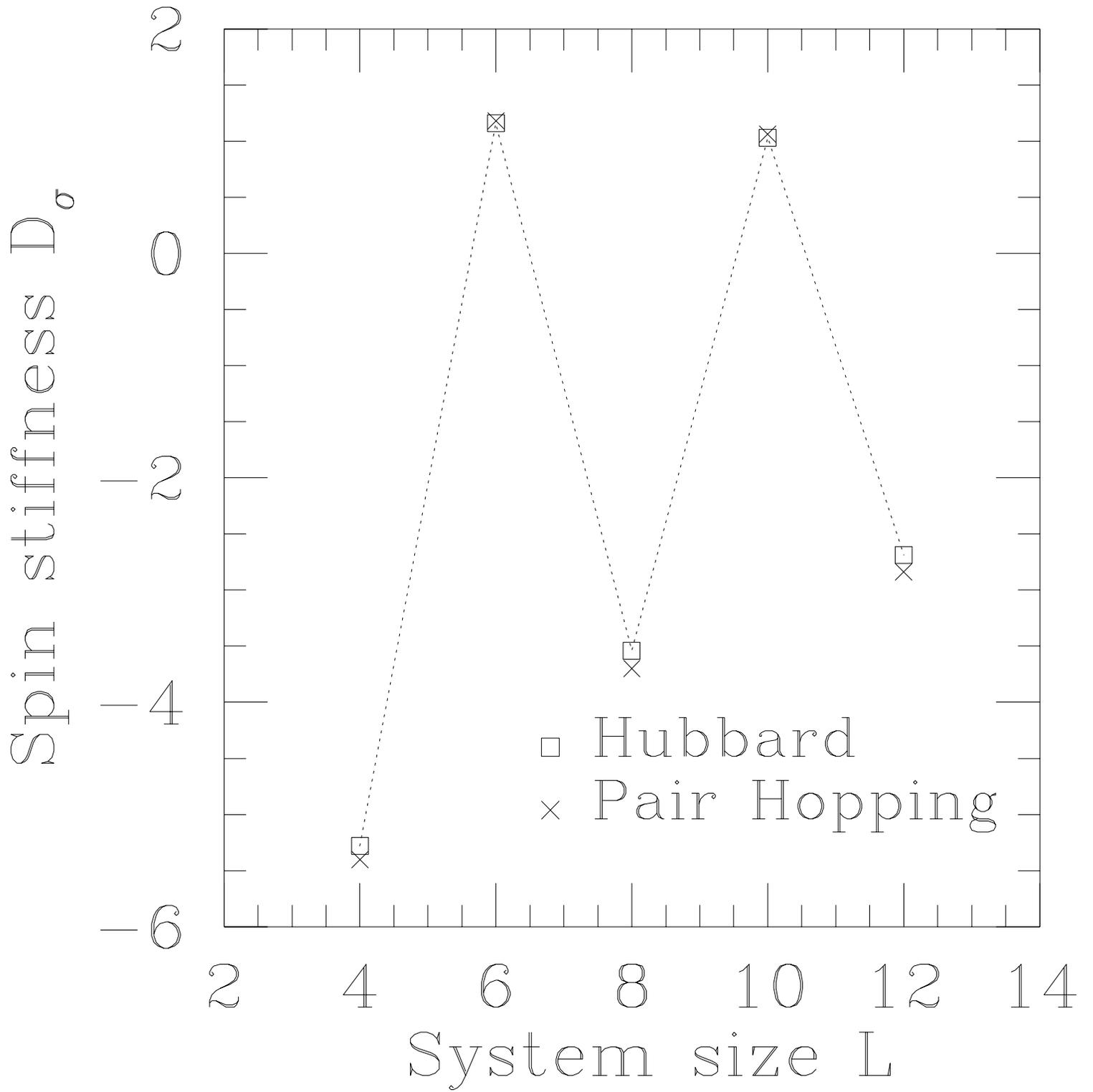
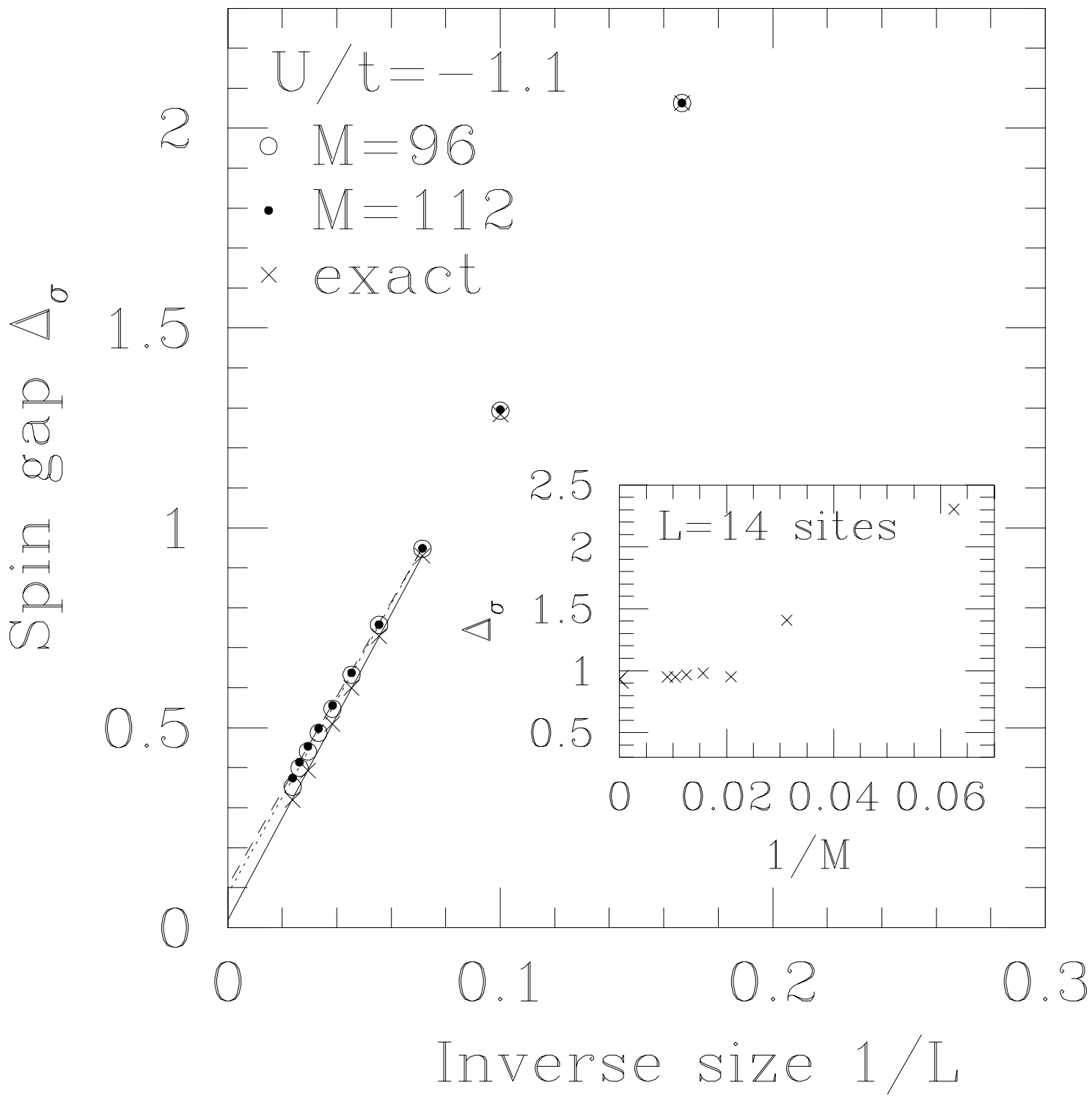
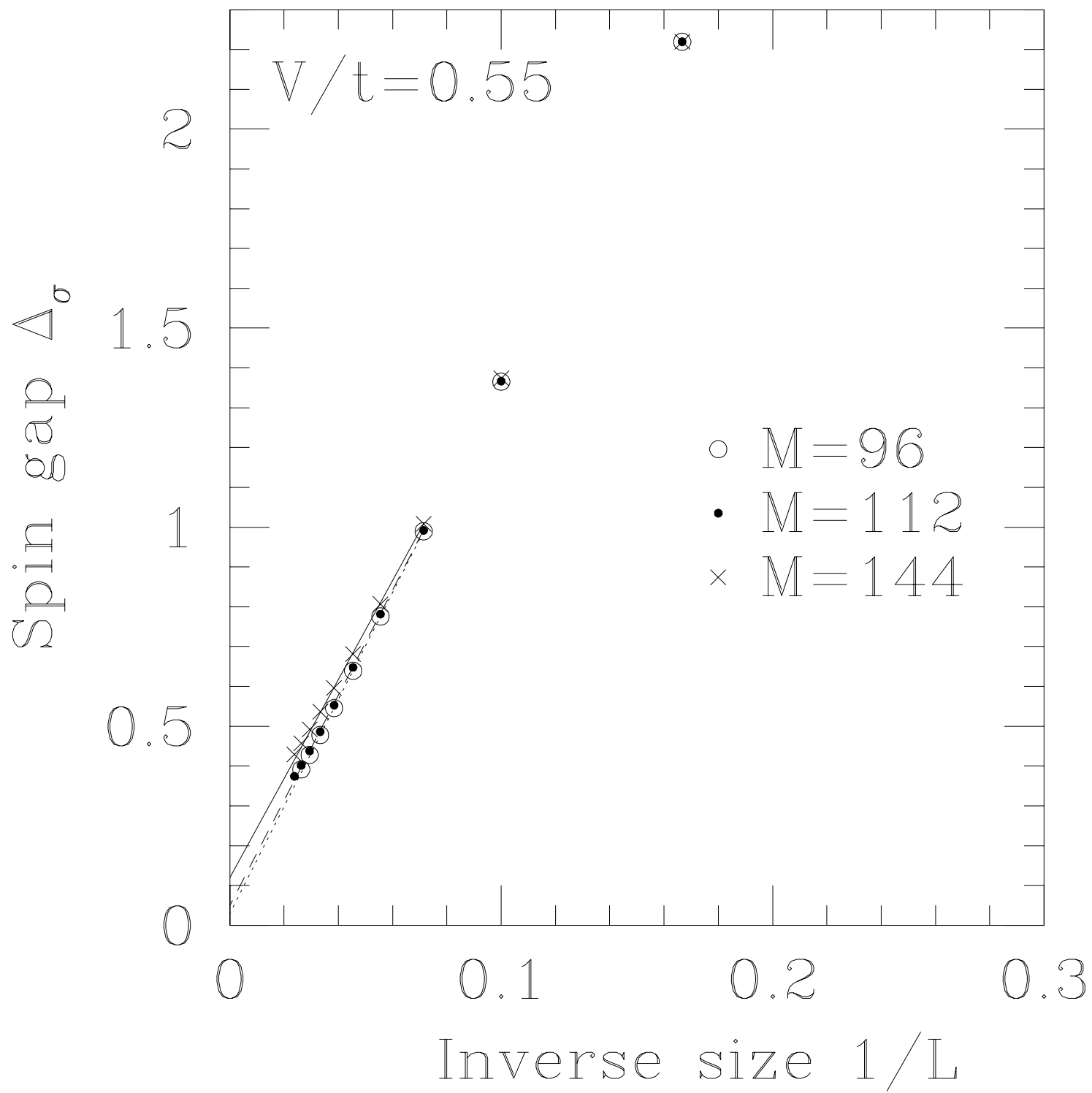
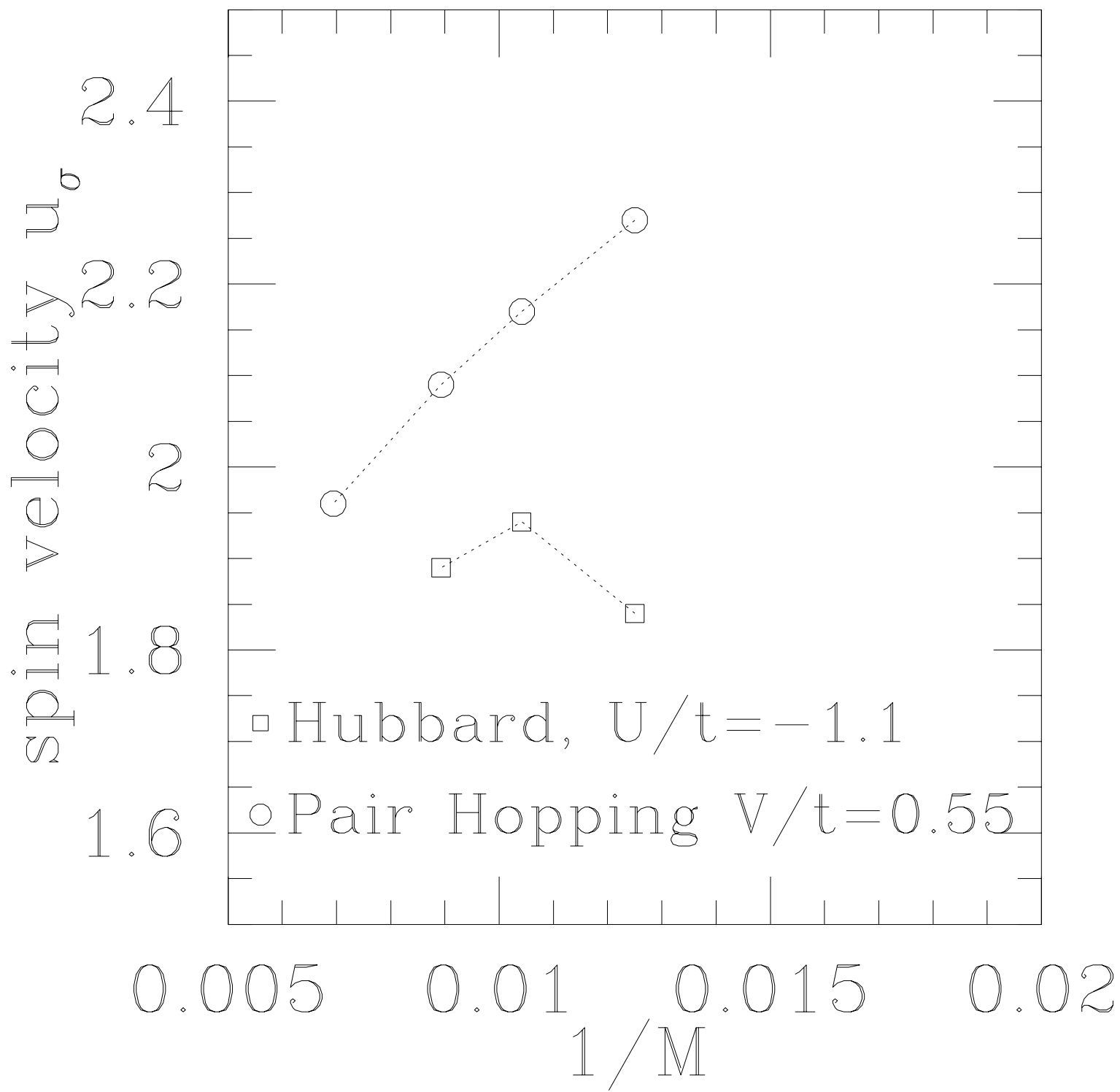
\end{document}